# Two General Methods for Population Pharmacokinetic Modeling: Non-Parametric Adaptive Grid and Non-Parametric Bayesian

*Tatiana Tatarinova*[*1], *Michael Neely*[*1], *Jay Bartroff*[1,2], *Michael van Guilder*[1], *Walter Yamada*[1,4], *David Bayard*[1,3], *Roger Jelliffe*[1], *Robert Leary*[1,5], *Alyona Chubatiuk*[2] and *Alan Schumitzky*[1,2]*

1 Laboratory of Applied Pharmacokinetics, Keck School of Medicine, University of Southern California, Los Angeles, CA, USA
2 Department of Mathematics, Dornsife College of Letters and Science University of Southern California, Los Angeles, CA, USA
3 Jet Propulsion Laboratory, California Institute of Technology, Mail Stop 198326, 4800 Oak Grove Drive, Pasadena, CA, USA
4 Department of Psychology, Azusa Pacific University, Azusa, CA, USA
5 Pharsight Corporation, Cary, NC, USA
* Joint first authors

Corresponding author email: tatiana.tatarinova@lapk.org

**ABSTRACT:**

Population pharmacokinetic (PK) modeling methods can be statistically classified as either parametric or nonparametric (NP). Each classification can be divided into maximum likelihood (ML) or Bayesian (B) approaches. In this paper we discuss the nonparametric case using both maximum likelihood and Bayesian approaches. We present two nonparametric methods for estimating the unknown joint population distribution of model parameter values in a pharmacokinetic/pharmacodynamic (PK/PD) dataset. The first method is the NP Adaptive Grid (NPAG). The second is the NP Bayesian (NPB) algorithm with a stick-breaking process to construct a Dirichlet prior. Our objective is to compare the performance of these two methods using a simulated PK/PD dataset. Our results showed excellent performance of NPAG and NPB in a realistically simulated PK study. This simulation allowed us to have benchmarks in the form of the true population parameters to compare with the estimates produced by the two methods, while incorporating challenges like unbalanced sample times and sample numbers as well as the ability to include the covariate of patient weight. We conclude that both NPML and NPB can be used in realistic PK/PD population analysis problems. The advantages of one versus the other are discussed in the paper. NPAG and NPB are implemented in R and freely available for download within the *Pmetrics* package from www.lapk.org.

**INTRODUCTION**

Population pharmacokinetic (PK) modeling involves estimating an unknown population distribution based on data from a collection of nonlinear models. One important application of this method is the analysis of clinical PK data. A drug is given to a population of subjects. In each subject, the drug's behavior is stochastically described by an unknown subject-specific parameter vector $\theta$. This vector $\theta$ varies significantly (often genetically) between subjects, which accounts for the variability of the drug response in the population. The mathematical problem is to determine the population parameter distribution $F(\theta)$ based on the clinical data.

The distribution $F$ determines the variability of the PK model over the population. From an estimate of this distribution, means and credibility intervals can be obtained for all moments of F and, more generally, for any functional of $F$, such as a target serum concentration after a given dosage regimen.

The importance of this problem is underscored by the FDA: "Knowledge of the relationship among concentration, response, and physiology is essential to the design of dosing strategies for rational therapeutics. Defining the optimum dosing strategy for a population, subgroup, or individual patient requires resolution of the variability issues." [1]

Population PK modeling approaches can be classified statistically as either parametric or nonparametric. Each can be divided into maximum likelihood or Bayesian methods. While we focus on the nonparametric approaches in this paper, for completeness we discuss all four approaches very briefly below.

The **parametric maximum likelihood** approach is the oldest and most traditional. One assumes that the parameters come from a known, specified probability distribution (the population distribution) with certain unknown population parameters (e.g. normal distribution with unknown mean vector $\mu$ and unknown covariance matrix $\Sigma$). The problem then is to estimate these unknown population parameters from a collection of individual subject data (the population data). The first and most widely used software for this approach has been the NONMEM program developed by Sheiner and Beal [2], [3]. There are other parametric maximum likelihood programs currently available, such as Monolix [4] and ADAPT [5]. The ADAPT software also allows for parametric mixtures of normal distributions, see [6] and [7]. Asymptotic confidence intervals can be obtained about these population parameters. Here "asymptotic" means as the number of subjects in the population becomes large.

The **nonparametric maximum likelihood (NPML)** approach was initially developed by Lindsay [8] and Mallet [9]. In contrast to parametric approaches, NPML makes no assumptions about the shapes of the underlying parameter distributions.  It directly estimates the entire joint distribution. This permits discovery of unanticipated, often genetically determined, non-normal and multimodal subpopulations, such as fast and slow metabolizers. The NPML approach is statistically consistent  [10]. This means that as the number of subjects gets large, the estimate of $F$ given the data converges to the true $F$. Consequently so are its continuous functionals such as means and covariances.  The main drawback of the NPML approach is that it is not easy to determine even asymptotic confidence intervals about parameter estimates. For example, bootstrap methods have been used   [11], but are extremely computationally intensive.

The **Bayesian** approaches are much newer. In the **parametric Bayesian** approach, one assumes that the population parameters (e.g. $(\mu, \Sigma)$ in the normal case)) are themselves random variables with known prior distributions. The problem then is to estimate the conditional distribution of the population parameters given the population data and the prior distributions. The most widely used approach is based on Monte Carlo Markov Chain (MCMC) methods.

Population PK analysis can be performed using the software packages WINBUGS  [12], and JAGS  [13].  Because this method is Bayesian, rigorous credibility intervals can be obtained for population parameters independent of the sample size. Of course, questions remain about convergence of the MCMC sampler and sensitivity to the prior assumptions.

The **nonparametric Bayesian** approach is much more flexible. One can assume that the population distribution $F$ is totally unknown and random with a Dirichlet process prior. This approach has only been applied to a few PK problems [14] [15], [16], [17]. A general purpose software package for population PK modeling has not been developed. This is one of the goals of the present paper.

## THE NONPARAMETRIC APPROACHES

We have developed two general nonparametric (NP) algorithms for estimating the unknown population distribution of model parameter values in a pharmacokinetic/pharmacodynamic (PK/PD) dataset [18], [19], [20].  The first method is the NP Adaptive Grid (NPAG) algorithm, which we have used in our USC Laboratory of Applied Pharmacokinetics for many years  [19]. This method calculates the maximum likelihood estimate of the population distribution with respect to *all* distributions.   Compared with most parametric population modeling methods, NPAG calculates exact, rather than approximate likelihoods, and it easily discovers unexpected sub-groups and outliers  [21], [22].

Since NPAG is an NPML method, it cannot easily calculate confidence intervals around parameter estimates. This motivated us to develop the second NP method described here. We used an NP Bayesian (NPB) algorithm with a stick-breaking process [23], to construct a Dirichlet process prior. More details are given below. The NPB algorithm provides a Bayesian estimate of this totally unknown population distribution, including rigorous (not asymptotic) credibility intervals around all parameter estimates for any sample size.

Both NPAG and NPB estimate the unknown population distribution as a discrete distribution. These discrete representations are perfectly suited for multiple-model adaptive control in which required integrals over the population distribution become finite sums [24]. By combining discrete distributions that are free of a priori assumptions on shape with credibility intervals, NPB combines the best of parametric and nonparametric methods.

The outline of the paper is as follows. First, we describe the mathematical and statistical details of the population PK/PD modeling problem. Then we describe the mathematical and statistical details of the NPAG and NPB algorithms. Next we give the results for our simulated PK/PD study data. The paper closes with conclusions and work for the future.

**THE POPULATION PK/PD MODEL**

Consider a sequence of experiments where each one consists of a dosage regimen and a set of measurements at several time points on one of $N$ individual subjects. The measurement model for the $i^{th}$ subject is:

(1) $\qquad Y_i = h_i(\theta_i) + e_i, \, i = 1,...,N$

where the vectors $Y_i$ are the observed measurements, e.g. serum concentrations, PD effects, etc. The components of the vector $\theta_i$ represent the unknown model and noise parameters defined on a space $\Theta$; $h_i(\theta_i)$ represents the noise-free output depending on the dosage regimen and the measurement schedule. The noise vectors $e_i$ are assumed to be independent, normal random variables with zero mean and covariance $\Sigma_i = \Sigma_i(\theta_i)$ which may depend on $\theta_i$.

The $\{\theta_i\}$ are independent and identically distributed with common (but unknown) probability distribution $F$. The population analysis problem is to estimate $F$ based on the data $Y^N = (Y_1,...Y_N)$.

**Algorithms**

The next two sections describe the mathematical and statistical details of the NPAG and NPB algorithms.

## NPAG ALGORITHM (NONPARAMETRIC ADAPTIVE GRID)

NPAG is an adaptive grid algorithm for finding the nonparametric maximum likelihood estimate of the population distribution. It was developed over a number of years at the Laboratory of Applied Pharmacokinetics, USC, by Alan Schumitzky [25], Robert Leary [26], and James Burke from the University of Washington, see also [27].

NPAG is based on a primal-dual interior point method. In this paper we present a brief outline of this approach. Consider Eq. (1). The log likelihood is given by

$$\log p(Y^N \mid F) = \log \prod_{i=1}^{N} \int p(Y_i \mid \theta_i) \, dF(\theta_i) \,.$$

The Maximum Likelihood distribution $F^{ML}$ maximizes $p(Y^N \mid F)$ over the space of all distributions $F$ defined on $\Theta$. Using Caratheodory's theorem and the results of Lindsay [8] and Mallet [9], it follows that $F^{ML}$ can be found in the class of discrete distributions with at most $N$ support points. In this case we write

(2)     $F^{ML}(\theta) = \sum_{k=1}^{K} w_k \delta_{\phi_k}(\theta)$,

where $\phi = (\phi_1, ...., \phi_K)$ are the support points of $F^{ML}$; $w = (w_1, ..., w_K)$ are the corresponding weights (probabilities) which sum to unity; and $\delta_{\phi}$ is the Dirac measure with mass 1.

Consequently, to maximize $\log p(Y^N \mid F)$, it is sufficient to maximize

(3)     $\log p(Y^N \mid w, \phi) = \sum_{j=1}^{N} \log \sum_{k=1}^{K} w_k \, p(Y_i \mid \phi_k), \ K \leq N$

with respect to the vectors $\{w_k\}$ and $\{\varphi_k\}$. If the support points $\{\phi_k\}$ are known, then maximization of $\log p(Y^N \mid w)$ with respect to the weights $\{w_k\}$ in Eq. (3) is a convex optimization problem and can be solved very efficiently.

The approach used in NPAG can be briefly described as follows: First solve the optimization problem for the weights related to Eq. (3) over a large but fixed grid $G_0$ of support points. Usually $G_0$ is taken to be a large number of so-called Faure points which optimally cover $\Theta$ [28]. Then reduce the grid $G_0$ by deleting points with very low probability to get a new grid $G_1$.

Then expand the grid $G_1$ by adjoining to each point $\phi_0$ in $G_1$ the vertices of a hypercube with $\phi_0$ as its center. This adds $2^{\dim \Theta}$ points to $G_1$ resulting in an expanded grid $G_2$. This cycle is repeated with $G_2$ replacing $G_0$. The process is continued until convergence is achieved.

*Optimization over fixed support points.*
The main part of the calculation comes in the optimization of the weights over a fixed grid of support points. Start with a set of support points $\phi$. Let $\Psi_{i,k} \equiv p(Y_i | \phi_k)$. Assume that the row sums of the $N \times K$ matrix $\Psi = [\Psi_{i,k}]$ are strictly positive (note that $\sum_{k=1}^{K} w_k p(Y_i | \phi_k) = (\Psi w)_i$). For any $K$-dimensional vector $z = (z_1, ..., z_K)^T$ we can define the function:

$$\Phi(z) = -\sum_{k=1}^{K} \log z_k, \text{ if } z > 0$$
$$= + \infty, \text{ otherwise}$$

Maximizing Eq. (3) with respect to the weights $\{w_k\}$ is equivalent to the solving the following two problems:

(P) Primal Problem: Minimize $\Phi(\Psi w)$ subject to $e^T w = 1$ and $w \geq 0$ where $e$ is the $K$-dimension column vector of with components all equal to one,

Now $grad[\Phi(\Psi w)] = grad[\Psi^T q]$, where $q = (q_1, ... q_n)^T$ and $q_i = \dfrac{1}{(\Psi w)_i}$. The Fenchel convex dual is then given by the Dual problem:

(D) Dual Problem: Minimize $\Phi(q)$ subject to $\Psi^T q \leq Ke$ and $q \geq 0$

*Duality Theorem and Karush-Kuhn-Tucker Conditions*
Solutions to (P) and (D) always exist, with the solution to (D) unique. Also $w$ solves (P) and $q$ solves (D) if and only if the Karush-Kuhn-Tucker [29] [30] conditions are satisfied:

(KKT) $\quad \Psi^T q + s - Ke = 0$
$\qquad\quad z - \Psi w = 0$
$\qquad\quad QZe - e = 0$
$\qquad\quad WSe = 0 \qquad\qquad$ ,

where $s$ is a non-negative $N$-dimensional vector (slack variables) and $Q=diag(q)$, $W=diag(w)$, $Z=diag(z)$, $S=diag(s)$. The primal-dual interior point method finds a solution to the above non-linear system of equations [27]. The Jacobian of the system is singular at the solution, so the strategy is to approach the solution along the central path $(w(\rho), q(\rho))$ as $\rho \downarrow 0$. In this case, the equation $WSe = \mathbf{O}$ is replaced by $WSe = \rho e$ and the (KKT) conditions become:

$$(\text{KKT}\rho) \quad \Psi^T q + s - Ke = 0$$
$$z - \Psi w = 0$$
$$QZe - e = 0$$
$$WSe - \rho = 0$$

The $(\text{KKT}\rho)$ equations are solved by Newton's method for a sequence of $\rho$ values tending to zero. The limit solution then solves both the primal and dual problems. The whole process converges quadratically and is very fast.

*Grid Adaptation: Reduction and Expansion*

As described earlier, the reduction of a current grid is accomplished by deleting support points with very low probability. The value of the likelihood function before and after grid reduction is essentially the same. The expansion of a current grid adds $2^{\dim\Theta}$ points to the grid. The optimization process over this new expanded grid can only increase the value of the likelihood function. When this increase is essentially zero, the whole process has converged. Exact details of solving the $(\text{KKT}\rho)$ equations and of grid adaptation will be published separately.

**NPB (NONPARAMETRIC BAYESIAN)**

There are two common ways to construct a Bayesian prior using a Dirichlet process: "marginal" and "full conditional" methods. In the framework of our Nonparametric Bayesian algorithms we implemented both approaches as described below. We now consider Eq. (1) from a Bayesian point of view. In this case the distribution $F$ is considered to be a random variable. The simplest prior distribution for $F$ is the so-called Dirichlet process, see [14] [17] for details. In this case we write $F \sim D(\alpha G_0)$ where the distribution $G_0$ will be our prior estimate of $F$, and where the number $\alpha$ will be the strength of that guess. As before, we assume $\theta_i \sim F$. Now the population analysis problem is to estimate the full conditional distribution of $F$ given the data $Y^N$. Most methods to solve this problem employ a marginal approach. However, $F$ can be "integrated out" leaving a much simpler problem for the $\theta_i$. The resulting model for the $\theta_i$ is given by the Polya Urn representation:

$$\theta_{n+1} \mid \theta_1,\dots,\theta_n \sim \frac{\alpha}{\alpha+n} G_0 + \frac{1}{\alpha+n} \sum_{j=1}^{n} \delta_{\theta_j}$$

The marginal approach leads to a Gibbs sampler algorithm for estimating $E[F \mid Data]$, i.e., the expected value of $F$ given the data, but not its distribution function [31]. This approach is commonly used. However, it does not solve the classical population analysis problem as stated above, for example, to estimate the full conditional distribution of $F$. To solve this problem we employ the Full Conditional Method. In this we estimate the full conditional distribution of $F$ given the data $Y^N$, not just the expected value of $F$.

The Full Conditional Method begins with a definition of the Dirichlet Process called the Stick-Breaking representation, see Sethuramam [32] and Ishwaran and James [23]. Consider the random distribution:

(5)     $$F(\theta) = \sum_{k=1}^{\infty} w_k \delta_{\phi_k}(\theta)$$

where the random vectors $\{\phi_k\}$ are independent and identically distributed (*iid*) from the known distribution $G_0$ and the weights $\{w_k\}$ are defined from the so-called stick-breaking process as follows:

$v_k \sim Beta(1,\alpha), k = 1,\dots,\infty;$

$w_1 = v_1, w_k = (1-v_1)(1-v_2)\dots(1-v_{k-1})v_k,$ for $k = 1,\dots,\infty,$

$w_1 = v_1, w_k = (1\text{-}v_1)(1\text{-}v_2)\dots(1\text{-}v_{k\text{-}1})v_k$

where $Beta(1,\alpha)$ is the Beta distribution with parameters $(1,\alpha)$. The name "Stick Breaking" comes from the fact that the $v_k$ are random cuts on a stick of length 1 and the $w_k$ are the lengths of the pieces broken off. This gives an informal proof that the $w_k$ sum to 1. It is shown in Sethuraman [32] that the random distribution $F \sim D(\alpha G_0)$ if and only if $F$ can be written in the form of Eq. (5).

Below we show how to use the Stick Breaking representation to estimate $F \mid Data$, not just $E[F \mid Data]$. The estimate of $F \mid Data$ leads to an estimate of all moments and their corresponding credibility intervals. More generally, the full conditional method can be used to estimate any functional of *F*, such as a target serum concentration profile to be achieved by a given dosage regimen, with its corresponding credibility interval.

**TRUNCATED STICK-BREAKING**

Ishwaran and James [23] consider replacing the infinite sum in Eq. (5) by a truncated version:

(6) $\qquad F_K(\theta) = \sum_{k=1}^{K} w_k \delta_{\phi_k}(\theta)$,

where it is now required that $v_K = 1$ so the series of weights sums to one. They show that the truncated version $F_K$ is virtually indistinguishable from $F$ for sufficiently large $K$. The only problem now is the size of $K$. Ishwaran and James [23] have suggested that $K=50$ is sufficient. In this paper we show that this number can be dramatically reduced.

Note that Eq. (6) has exactly the same form as the Eq. (2) for $F^{ML}$. The difference is that in Eq. (2), the weights $\{w_k\}$ and support points $\{\phi_k\}$ are deterministic, while in Eq. (6) the same quantities are random. In other words, $F^{ML}$ is a deterministic distribution while $F_K$ is a random distribution.

**Full Conditional Approach**. Let us assume now that we have a sufficient number of components in Eq. (6) to approximate the infinite sum in Eq. (5) accurately. Using Eqs. (3,5,6) we have the new model given by

(7) $\qquad Y_i = h_i(\theta_i) + e_i; \quad \theta_i \sim F_K; \quad F_K(\theta) = \sum_{k=1}^{K} w_k \delta_{\phi_k}(\theta)$

where the random vectors are *iid* from the distribution $G_0$ and the weights are defined by Eq. (5b) with $v_K = 1$.

Eq. (7) defines a mixture model with an unknown but finite number of components. Much is known about this subject [33], [34], [35]. For a fixed number of components $K$, the posterior distribution of the weights $\{w_k\}$ and the support points $\{\phi_k\}$ can be determined by the Blocked Gibbs Sampling [23]. Consequently, for a fixed $K$, the posterior estimates of the support points $\{\phi_k\}$ and the weights $\{w_k\}$ are straightforward to calculate. As opposed to the Gibbs Sampler for the Marginal Method, the Gibbs Sampler for the NPB approach directly involves the distribution $F_K$.

**Sampling the posterior of $F_K$, $E[F_K]$ and moments of $F_K$.**

Let be $w_k^m, \phi_k^m, \ k=1,...,K; \ m=1,....M$ samples of $w_k, \phi_k$ from the Gibbs Sampler after the sampler has "converged". Then we have:

<u>Samples from</u> $F_K \mid Y^N$ : $\quad F_K^{(m)}(\theta) = \sum_{k=1}^{K} w_k^m \delta_{\phi_k^m}(\theta), \ m=1,...,M$

<u>Samples from the moments of</u> $F*|Y^N$: Let $\mu_j = \int \theta^j dF(\theta)$ be the j$^{th}$ moment of $F_K | Y^N$.

Then samples of $\mu_j$ are given by: $\mu_j^{(m)} \approx \sum_{k=1}^{K} w_k^m (\phi_k^j)^m$, $m = 1, ..., M$.

In particular, samples from the first moment or mean are given by

$$\mu^{(m)} \equiv \mu_j^{(m)} = \int \theta dF^{(m)}(\theta) = \sum_{k=1}^{K} w_k^m \phi_k^m, \; m = 1, ..., M.$$

A histogram of the values of $\mu^{(m)}$ gives the estimated distribution of $\mu = \int \theta dF(\theta)$, and the Bayesian credibility intervals are derived from it.

<u>Samples from the expected value</u> $E[F_K | Y^N]$. For this quantity we calculate:

$$E[F_K | Y] \approx (1/M) \sum_{m=1}^{M} F_K^{(m)}(\theta).$$

To assess the performance of our algorithm, we can compare our estimate of $E[F_K | Y^N]$ with the estimate from the marginalization method.

<u>Choice of K.</u> We have implemented the Gibbs sampler from Ishwaran and James [23], Sec. 5.2, using the software package JAGS [13]. An important feature of this algorithm is that it keeps track of the number K* of distinct components in the K component mixture. If K is chosen too small, the algorithm will alert the user by indicating that K*= K. See [36], [37] for applications to pharmacokinetics using truncated stick-breaking methods.

A more sophisticated way of choosing K is based on new results for Retrospective Sampling [31] and Slice Sampling [38], [39]. In these methods the infinite sum in the stick-breaking representation of Eq. (5) is retained but only as many terms in the sum are used as are needed in the calculation.

**COMPARISION OF NPAG AND NPB METHODS**

Both NPAG and NPB estimate the entire probability distribution $F$ of PK/PD parameters in a population modeling setting. NPAG is a deterministic method using the convexity of the resulting nonparametric maximum likelihood problem. The optimization algorithm in NPAG is based on "state of the art" primal-dual interior-point theory. It has been used in our laboratory for many years and can handle PK/PD problems defined by 20-30 differential and algebraic equations containing 20-30 unknown parameters. The algorithm is very stable and fast. It always determines an optimal solution to the problem. The only drawback with NPAG is it does not directly determine confidence intervals of the parameters of interest. (When the number of subjects in the population is large, then the asymptotic confidence intervals can be obtained with additional computing by bootstrap methods.)

NPB is a stochastic Monte Carlo Markov Chain (MCMC) method. The unknown probability distribution $F$ is considered to be a random variable with a Dirichlet process prior. The Dirichlet process is implemented with the Sethuraman's stick-breaking representation. The algorithm used to estimate $F$ is a Metropolis-within-Gibbs sampling (GS) scheme. For the example in this paper, the program JAGS is used to implement this scheme. This implementation of GS is composed of three parts: First a number of samples of GS are burned to remove dependence on the initial conditions; then GS is run for a large number of iterations until "convergence" is achieved. Then after convergence, GS is run some more to get the samples used for the actual estimation and plotting of results. The number of samples required for convergence is a delicate issue. There are many candidates to test convergence of MCMC algorithms. No one method is perfect. We use the Gelman-Rubin method of parallel chains to determine convergence. Finally, being a Bayesian method, NPB can provide rigorous credibility intervals for any function of interest of the PK/PD parameters. These credibility intervals are accurate in both large and small population sample sizes.

In conclusion, for a given set of initial conditions, NPAG will always give the same results, whereas NPB may possibly give different results depending on the sampling scheme. On the other hand, no confidence intervals are available with NPAG (without asymptotic bootstrap), while rigorous Bayesian credibility intervals are defined for NPB no matter what the sample size. Consequently, it is extremely useful to be able to run both NPAG and NPB side by side and compare the results (as shown in this paper).

Finally, both NPAG and NPB estimate $F$ with a finite discrete probability distribution (as described in the paper). This result is vital for our resulting maximally precise "multiple model" dosage optimization and experimental design programs.

**EXAMPLE: A POPULATION PK STUDY**

We took dosing, sample times, and body weights from *N=35* infants enrolled in an IV zidovudine PK study as a template to simulate new observations after a short intravenous (IV) infusion of a hypothetical drug into a one compartment model with simulated PK parameter values. This provided a realistic simulated dataset with unbalanced doses, number of samples numbers and sample times in the population, but with known PK parameter values for each subject. We used the Monte Carlo simulator function in our R package Pmetrics [21]. In short, Pmetrics is our freely available R package for non-parametric and parametric population modeling and simulation. It can model multiple inputs and outputs simultaneously, with complex dosing regimens, inclusion of covariates, lag times, and non-zero initial conditions all available to the user. Specification of a model, based on algebraic equations or differential equations and incorporating any function of parameters and covariates, is done with a very simple text file. Detailed examples and model files can be found at http://www.lapk.org.

**Simulation model**. For the PK parameter values, we set the elimination rate constant $(Kel = \kappa)$ as a mixture of two normal distributions with arbitrary means of 0.5 and 1.0 1/h and weights of 0.3 and 0.7. The population average was equal to 0.85 1/h and located in the "valley" between the two modes. These parameter values produced realistic time-concentration profiles. The coefficients of variation (CV) for each distribution were set at 25%. We set the volume of distribution to be a single normal distribution, with a mean of 2.0 L/kg and standard deviation of 0.5 L/kg. The measurement noise, as a normal distribution with mean 0 and standard deviation $\sigma_e$=0.01, was added to each simulated observation.

Hence, we consider a one compartment model with *T=5 or 6* serum measurements (specific for each patient) for a population of *N=35* subjects. In this case $\Theta = (\kappa; V)$; where $\kappa$ is bimodal and $V$ is unimodal. Therefore the model that was used to simulate the data is described by the following equations:

$$Y_{ij} = \frac{R_i}{Wt_i V_i \kappa_i} \, (1 - e^{-d_i \kappa_i}) \, e^{-(t_{ij} - d_i)\kappa_i} + e_{ij}, \quad i = 1, ..., N; \quad j = 1, ..., n_i$$

$$e_{ij} \sim N(0, 0.01)$$

$$\kappa_i \sim 0.3 N(0.5, 0.125) + 0.7 N(1.0, 0.25)$$

$$V_i \sim N(2.0, 0.5)$$

where $R_i$ is the subject-specific infusion rate with $d_i$ infusion duration for zidovudine; $Wt_i$ is the body weight in kg for each subject; $t_{ij}$ is the time of the $j^{th}$ sample from subject $i$; and $e_{ij}$ is the measurement noise of the $j^{th}$ measurement noise in subject $i$. Values of $R_i$, $Wt_i$, $d_i$ were the defined patient-specific parameters in the original population of infants. The symbol ~ means "distributed as". To avoid negative parameter values we also set: $\kappa_i = |\kappa_i|$ and $V_i = |V_i|$.

**Estimation Model**. The NPB model used to analyze the data came from the stick-breaking representation, with *K=17*, see Eq. (6). This number of stick breaks (support points) for the NPB prior was based on the number of clusters created by the NPB algorithm and was determined by a manual iterative approach. If more than 17 support points were used the resulting probabilities assigned to additional support points were negligible. The base measure $G_0(\kappa, V)$ was given by:

$$\kappa \sim N(\kappa_0, 1/\tau_0^\kappa), \kappa_0 \sim N(0.8, 0.5), \tau_0^\kappa \sim \text{Gamma}(1,1)$$

$$V \sim N(V_0, 1/\tau_0^V), V_0 \sim N(2.0, 0.5), \tau_0^V \sim \text{Gamma}(1,1)$$

where *Gamma(a,b)* is the gamma distribution with parameters *a,b*. These are the common distributions traditionally chosen for means and variances. The user of the program can also make other choices.

Using the NPAG algorithm [19] from the Pmetrics software package [21], we calculated the maximum likelihood distribution $F^{ML}$, see Figure 2. Implementing the stick-breaking algorithm using the Rjags package [40], we calculated the estimated conditional distribution $F^{ML}$, shown in Figure 2. For the NPB algorithm, we used one Markov Chain for the Monte Carlo simulation, drawing every 10th posterior sample from iteration 10,000 to 10,500.

As a further comparison, we also fitted the data with the NONMEM FOCE algorithm, with V and K modeled as univariate normal distributions with an additive measurement error $e_{ij}$.

**Results**

Simulated observations with realistic, unbalanced sampling times and sample numbers ranged from <0.01 to 1.64 mg/L, calculated up to 8 hours after dosing, with 5–6 samples per subject at times that varied throughout the population, and which corresponded to the times that real infants in the source population had been sampled. *Figure 1* shows the simulated time-series. The whole NPAG optimization, including post-processing and report generation, took 18 seconds on a MacBook Pro with 2.54 GHz Intel core 2 Duo processors and 4 GB of RAM. On the same computer, NPB took 2 minutes.

Summaries of the simulated (True) values for KEL and VOL and of the weighted support points fitted by the NPB, NPAG and FOCE algorithms are shown in the Table 1. Figures 2-5 show the output of the NPB algorithm. In the Figures 2 and 3, the NPB estimates for Vol and Kel are plotted against the histogram of simulated values for volume of distribution. Figures 4 and 5 show NPB error in true - fitted parameter values for comparison of NPB estimated vs. the simulated values for volume of distribution and elimination constant. Figures 6 and 7 show the NPAG estimates for Vol and Kel compared against the histogram of simulated values for volume of distribution.

For 35 subjects NPB estimates values Vol and Kel for individual patients, but the estimated parameter distribution functions have too many peaks as compared to the "true" parameter distributions. When we increase the number of subjects to 70 or more (data not shown), the estimated parameter distribution functions has two modes for *Kel* and one mode for *Vol*, and parameter distributions approach the true population distributions. However, due to the nature of our simulation study (infant HIV patients enrolled in an IV zidovudine PK study) it is not realistic to expect large cohorts in a clinical setting.

## DISCUSSION

Bayesian methods are rapidly gaining recognition and popularity. A comprehensive overview of the general philosophy of Bayesian methods can be found in the book "The Bayesian Choice" [41]. Computational issues of MCMC methods are well described in "Monte Carlo Statistical Methods" [42]. Biostatistics applications are described in the chapter "Nonparametric Bayes Applications to Biostatistics" [43].  To the best of our knowledge, currently there are no textbooks that primarily discuss nonparametric pharmacokinetic modeling. However, our references [14], [15], [16], [17], [18] provide a good survey of this subject.

We have described two novel methods, NPAG and NPB, to estimate the population distribution F of PK parameters, have shown their excellent performance in a realistically simulated PK study. We also compared their performance to NONMEM, the widely-used FOCE algorithm. In this simple model, FOCE, a parametric method, was able to find the same the mean parameter values and standard deviations, but could not find the true non-normal distribution for K without resorting to post-hoc estimates.  In contrast, both NPAG and NPB are able to directly estimate the true distribution.  In future work beyond our proof-of-principle work on NPB here, we will show that more "challenging" data with greater noise and non-normal parameter value distributions are even better fitted by optimized non-parametric methods, i.e. NPAG and NPB. In this paper, our simulation allowed us to have benchmarks in the form of the true population parameter values to compare with the estimated values, while incorporating challenges like unbalanced sample times and sample numbers as well as the ability to include the covariate of patient weight.

The statistical problem of estimating $F$ has a direct utility in the form of individualized therapy of future patients because the estimate of $F$ can be used quickly and accurately to isolate a new patient's characteristics (i.e., parameters) and use this knowledge to tailor patient-specific efficacious treatment.  The NPB method is very flexible and has been used in many areas of applied statistics and bioinformatics outside PK, discussed, for example,  in [43].

NPAG and NPB represent two ends of the spectrum spanning frequentist (NPAG) to Bayesian (NPB) methodologies; they estimate the entire distribution $F$, not just parameter values. The two methods are the state-of-the-art in nonparametric population modeling, and they accurately estimate the parameter distributions without resorting to any a priori assumptions about the underlying form of these distributions. While NPAG is significantly faster at present, the main advantage of the NPB method is that it naturally allows for robust credibility intervals for all parameter estimates.

The simulation study presented above is performed in the setting of a real zidovudine trial which allows us to have benchmarks in the form of the true population parameters to compare with the estimates produced by the two methods, while incorporating realistic challenges like unbalanced sample times and sample numbers as well as the ability to include the covariate of patient weight. Figures 2-7 and Table 1 show that both methods focus on the marginal distributions of *Kel* (elimination rate constant) and *Vol* (volume of distribution) in our example and produce accurate estimates of their moments. We have previously shown that NPAG, as implemented in our Pmetrics R package, can directly and accurately detect true non-normal parameter distributions and outliers in an idealized simulated population [21]. In addition to confirming this property of NPAG with a more realistic study design here, we extend this property to our NPB algorithm.

Future refinements of the NPB algorithm include exploring convergence criteria, implementation of the Retrospective and Slice sampling methods to determine the correct number of stick breaks (i.e. support points, as opposed to the empiric approach described here), and generalization to even more complex PK models, including arbitrary models defined by ordinary differential equations. The software used to implement NPAG and NPB can be obtained from http://lapk.org/software.php.

**ACKNOWLEDGEMENTS**

Support from NIH: GM068968, EB005803, EB001978, NIH-NICHD: HD070996 and Royal Society: TG103083 is gratefully acknowledged.

# References

1. FDA. **FDA Guidance for Industry: Population Pharmacokinetics**. . 1999.

2. BEAL, S.; SHEINER, L. Estimating population kinetics. **Crit Rev Biomed Eng.** , v. 8, n. 3, p. 95-222, 1982.

3. BEAL, S. L.; SHEINER, L. B. NONMEM User's Guide. In:  **Nonlinear Mixed Effects Models for Repeated Measures**. University of California: San Francisco, 1992.

4. LAVIELLE, M.; MENTRÉ, F. Estimation of population pharmacokinetic parameters of saquinavir in HIV patients with the MONOLIX software. **Journal of Pharmacokinetics and Pharmacodynamics** , v. 34, n. 2, p. 229-49, 2007.

5. D'ARGENIO, D. Z.; SCHUMITZKY, A.; WANG, X. **ADAPT 5 User's Guide:Pharmacokinetic/Pharmacodynamic Systems Analysis Software.**. Los Angeles. 2009.

6. WANG, A.; SCHUMITZKY, A.; DARGENIO, D. Nonlinear random effects mixture models: Maximum likelihood estimation via the EM algorithm. **Comput. Stat. Data Anal.**, v. 51, p. 6614-6623, 2007.

7. WANG, A.; SCHUMITZKY, A.; DARGENIO, D. Population pharmacokinetic/pharmacodyanamic mixture models via maximum a posteriori estimation. **Comput. Stat. Data Anal.** , 2009.

8. LINDSAY, B. The Geometry of Mixture Likelihoods: A General Theory.. **Ann Stat** , v. 11, p. 86–94, 1983.

9. MALLET, A. A maximum likelihood estimation method for random coefficient regression models. **Biometrika**, v. 73, p. 645–656, 1986.

10. KIEFER, J.; WOLFOWITZ, J. Consistency of the Maximum Likelihood Estimator in the Presence of Infinitely Many Incidental Parameters. **The Annals of Mathematical Statistics.**, v. 27, n. 4, p. 887-906, 1956.

11. BAVEREL, P.; SAVIC, R.; KARLSSON, M. Two bootstrapping routines for obtaining imprecision estimates for nonparametric parameter distributions in nonlinear mixed effects models. **J Pharmacokinet Pharmacodyn.**, v. 38, n. 1, p. 63-82, Feb 2011. Epub 2010 Nov 13.

12. SPIEGELHALTER, D. J.; THOMAS, A.; BEST, N. G. **WinBUGS Version 1.4 User Manual, MRC Biostatistics Unit.**  2004.

13. PLUMMER, M. **JAGS:** A Program for Analysis of Bayesian Graphical Models Using Gibbs Sampling. Proceedings of the 3rd International Workshop on Distributed Statistical Computing (DSC 2003). Vienna, Austria. 2003.

14. WAKEFIELD, J.; WALKER, S. Bayesian nonparametric population models: formulation and comparison with likelihood approaches.. **J Pharmacokinet Biopharm**, v. 25, p. 235–253, 1997.

15. WAKEFIELD, J.; WALKER, S. Population models with a nonparametric random coefficient distribution. **Sankhya Series B**, v. 60, p. 196–212, 1998.

16. MUELLER, P.; ROSNER, G. A Bayesian Population Model With Hierarchical Mixture Priors Applied to Blood Count Data. **J Am Stat Assoc** , v. 92, p. 1279–1292, 1997.

17. ROSNER, G.; MUELLER, P. Bayesian population pharmacokinetic and pharmacodynamic analyses using mixture models. **J Pharmacokinet Biopharm**, v. 25, p. 209–233, 1997.

18. WANG, J. Dirichlet Processes in Nonlinear Mixed Effects Models. **Communications in Statistics: Simulation and Computation** , v. 39, p. 539-556, 2010.

19. YAMADA, Y. et al. **The Nonparametric Adaptive Grid Algorithm for Population Pharmacokinetic Modeling.** USC 2012.

20. NEELY, M. et al. **Non-Parametric Bayesian Fitting:** A Novel Approach to Population Pharmacokinetic Modeling. Poster presented at: Population Analysis Group in Europe. Venice, Italy, 2012.

21. NEELY, M. et al. Accurate Detection of Outliers and Subpopulations With Pmetrics, a Nonparametric and Parametric Pharmacometric Modeling and Simulation Package for R. **Theraputic Drug Monitoring**, v. 34, n. 4, p. 467–476, 2012.

22. BUSTAD, A. et al. Parametric and Nonparametric Population Methods: Their Comparative Performance in Analysing a Clinical Dataset and Two Monte Carlo Simulation Studies. **Clinical Pharmacokinetics**, v. 45, n. 4, p. 365-383, 2006.

23. ISHWARAN, H.; JAMES, L. Gibbs Sampling Methods for Stick-Breaking Priors. **J Am Stat Assoc** , v. 96, p. 161–173, 2001.

24. JELLIFFE, R. et al. Achieving target goals most precisely using nonparametric compartmental models and "multiple model" design of dosage regimens. **Ther. Drug. Monit.**, v. 22, p. 346–353, 2000.

25. SCHUMITZKY, A. Nonparametric EM Algorithms for Estimating Prior Distributions. **Applied Math and Computation**, v. 45, p. 141–157, 1991.

26. LEARY, R. et al. **An adaptive grid non-parametric approach to population pharmacokinetic/dynamic**

**(PK/PD) population models**. Proceedings, 14th IEEE symposium on Computer Based Medical Systems.  2001. p. 389–394.

27. BAEK, Y. **An Interior Point Approach to Constrained Nonparametric Mixture Models**. University of Washington. 2006.

28. FOX, B. L. Algorithm 647: Implementation and Relative Efficiency of Quasirandom Sequence Generators. **Transactions on Mathematical Software**, v. 12, n. 4, p. 362-376, 1986.

29. KARUSH, W. **Minima of Functions of Several Variables with Inequalities as Side Constraints**. University of Chicago. Chicago, Il. 1939.

30. KUHN, H.; TUCKER, A. **Nonlinear Programming**. Proceedings of the 2nd Berkeley Symposium.  [s.n.]. 1951. p. 481-492.

31. PAPASPILIOPOULOS, O.; ROBERTS, G. O. Retrospective Markov Chain Monte Carlo methods for Dirichlet process hierarchical models. **Biometrika**, v. 95, n. 1, p. 169-186, 2008.

32. SETHURAMAN, J. A constructive definition of Dirichlet priors. **Statistica Sinica**, v. 4, p. 639–650, 1994.

33. TATARINOVA, T. **Bayesian Analysis of Linear and Nonlinear Mixture Models**. USC. Los Angeles. 2006.

34. TATARINOVA, T.; BOUCK, J.; SCHUMITZKY, A. Kullback-Leibler Markov chain Monte Carlo--a new algorithm for finite mixture analysis and its application to gene expression data. **J Bioinform Comput Biol.**, v. 6, n. 4, p. 727-46, Ag 2008.

35. FRÜHWIRTH-SCHNATTER, S. **Finite Mixture and Markov Switching Models**. 1st ed. ed. New York: Springer, 2010. Springer Series in Statistics.

36. GHOSH, P.; ROSNER, G. A Semiparametric Bayesian Approach to Average Bioequivalence. **Statistics in Medicine**, v. 26, p. 1224-1236, 2007.

37. OHLSSEN, D.; SHARPLES, L.; SPIEGELHALTER, D. 'Flexible random-effects models using Bayesian semi-parametric models: applications to institutional comparisons. **Statistics in Medicine**, v. 26, p. 2088–2112, 2007.

38. WALKER, S. Sampling the Dirichlet mixture model with slices. **Communications in Statistics: Simulation and Computation**, v. 36, p. 45–54, 2007.

39. KALLI, M.; GRIFFEN, J.; WALKER, S. Slice Sampling Mixture Models. **Statistics and Computing**, v. 21, p. 93-105, 2011.

40. PLUMMER, M. **rjags: Bayesian graphical models using MCMC.** 2011.

41. ROBERT, C. **The Bayesian Choice**. 2nd. ed.  Springer, 2007.

42. ROBERT, C.; CASELLA, G. **Monte Carlo Statistical Methods**. 2nd. ed.  Springer, 2004.

43. DUNSON, D. Nonparametric Bayes Applications to Biostatistics. In: HJORT, N., et al. **Bayesian Nonparametrics**.  Cambridge University Press, 2010. p. 223-268.

## List of Tables

Table 1: Summaries of the simulated (True) values for elimination rate constant (KEL) and volume of distribution (VOL) and the corresponding values fitted by the Non-Parametric Adaptive Grid (NPAG), Non-Parametric Bayesian (NPB) and NONMEM algorithms.

| | True | | Estimated | | |
| --- | --- | --- | --- | --- | --- |
| | Distribution | Sample (Simulated) | NPAG fitted | NPB fitted | NONMEM fitted |
| Size | | 35 (subjects) | 23 (SP) | 17 (SP) | NA |
| KEL (1/h) | | | 0.77 (NA) | 0.76 (0.73–0.79) | 0.77(0.68-0.86) |
| mean (95% CI) | 0.85 (NA) | 0.77 (NA) | 0.27 (NA) | 0.24 (0.20–0.32) | 0.28(0.20-0.33) |
| SD (95% CI) | 0.22 (NA) | 0.27 (NA) | | | |
| VOL (L/kg) | | | 2.03 (NA) | 1.98 (1.92–2.03) | 2.03(1.93-2.12) |
| mean (95% CI) | 2.0 (NA) | 2.03 (NA) | 0.27 (NA) | 0.30 (0.25–0.40) | 0.28(0.21-0.34) |

# List of Figures

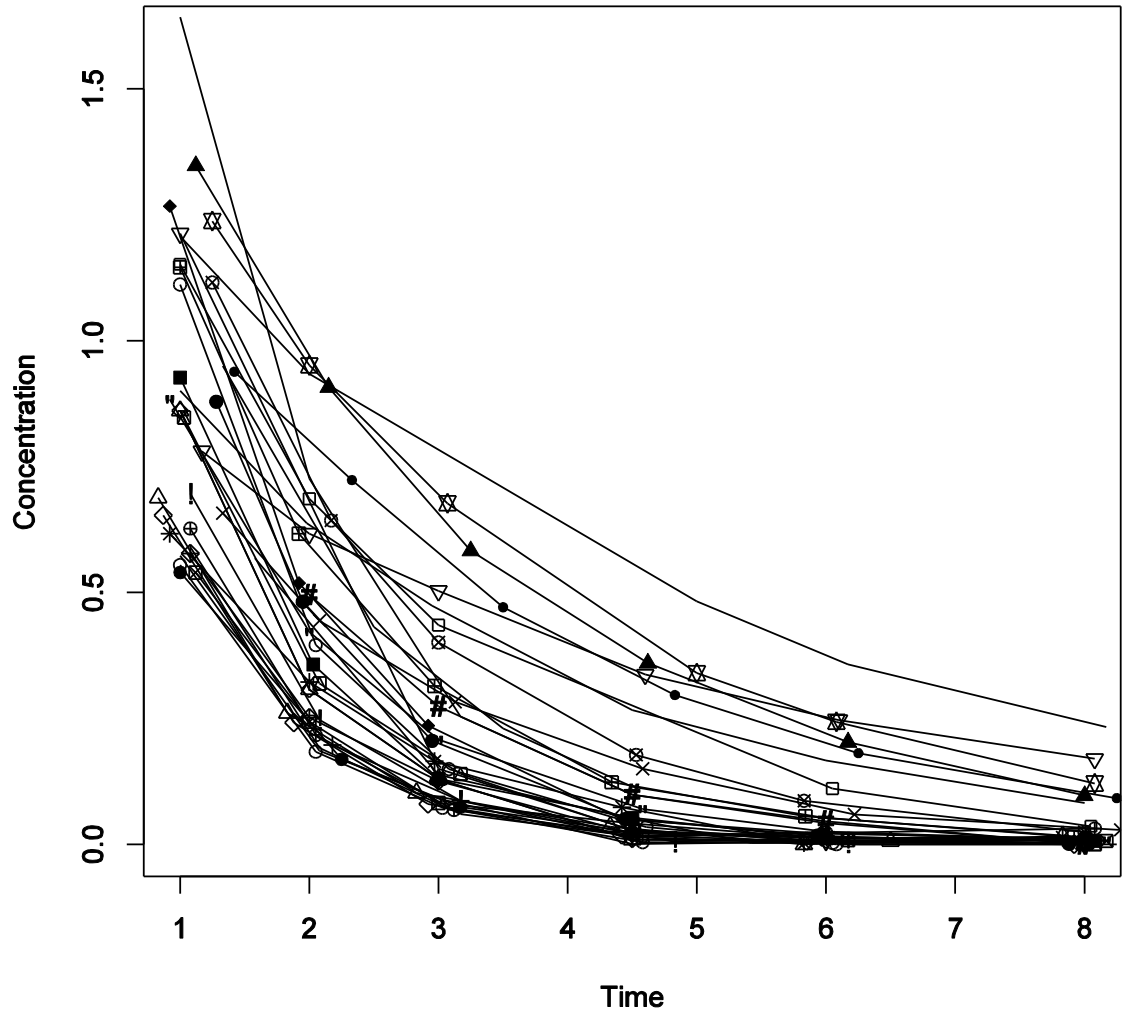

Figure 2: NPB distribution and simulated values for volume of distribution (VOL). Marginal distributions for simulated (true) parameter values are shown in black circles and seven filled histograms. The posterior distribution is represented in two ways 1) dark grey bars with binning (nbins=50) and 2) a smoothed sum of normal distributions about the means of the distributions for each of the 17 support points (solid black line). True population distribution is shown as a dashed line.

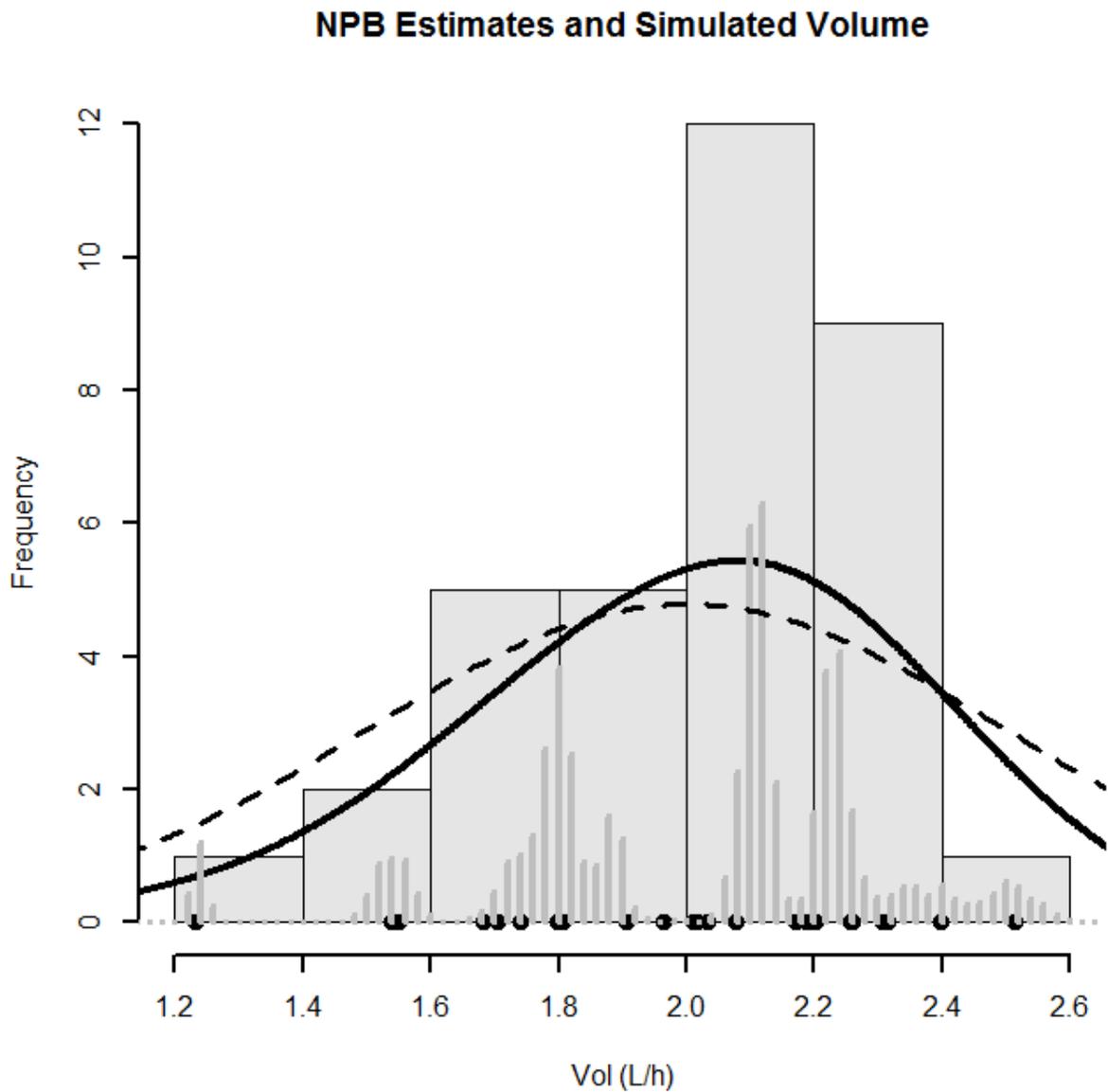

Figure 3: NPB distribution and simulated values for elimination rate constant (KEL). Marginal distributions for simulated (true) parameter values are shown in black circles and filled histograms. The posterior distribution is represented in two ways 1) dark grey bars with binning (nbins=50) and 2) a smoothed sum of normal distributions about the means of the distributions for each of the 17 support points (solid black line). True population distribution is shown as a dashed line.

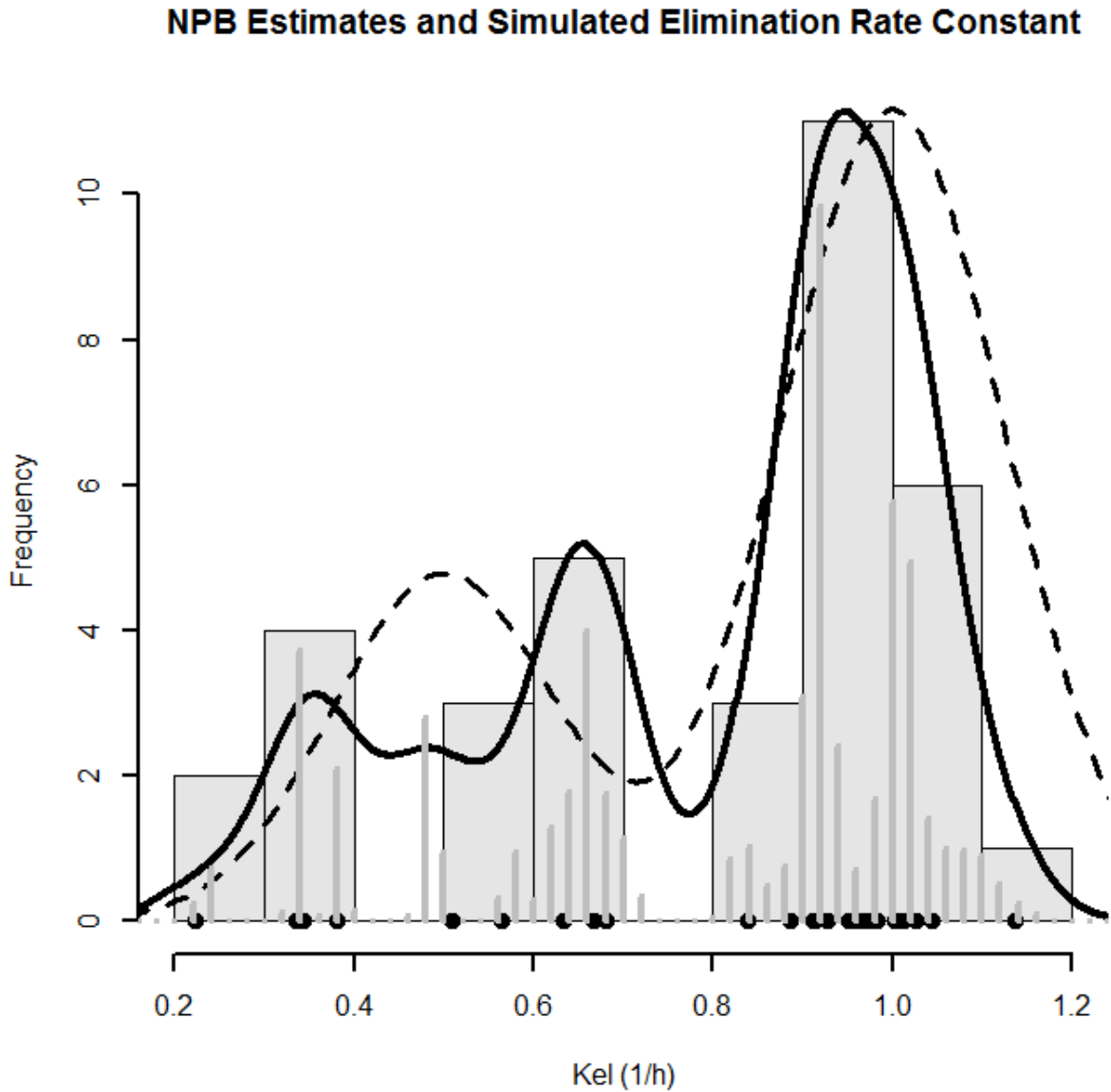

Figure 4: NPB error in true - fitted parameter values (VOL). Linear regression of simulated volume of distribution vs. predicted volume of distribution for each of the 35 simulated subjects using NPB algorithm.

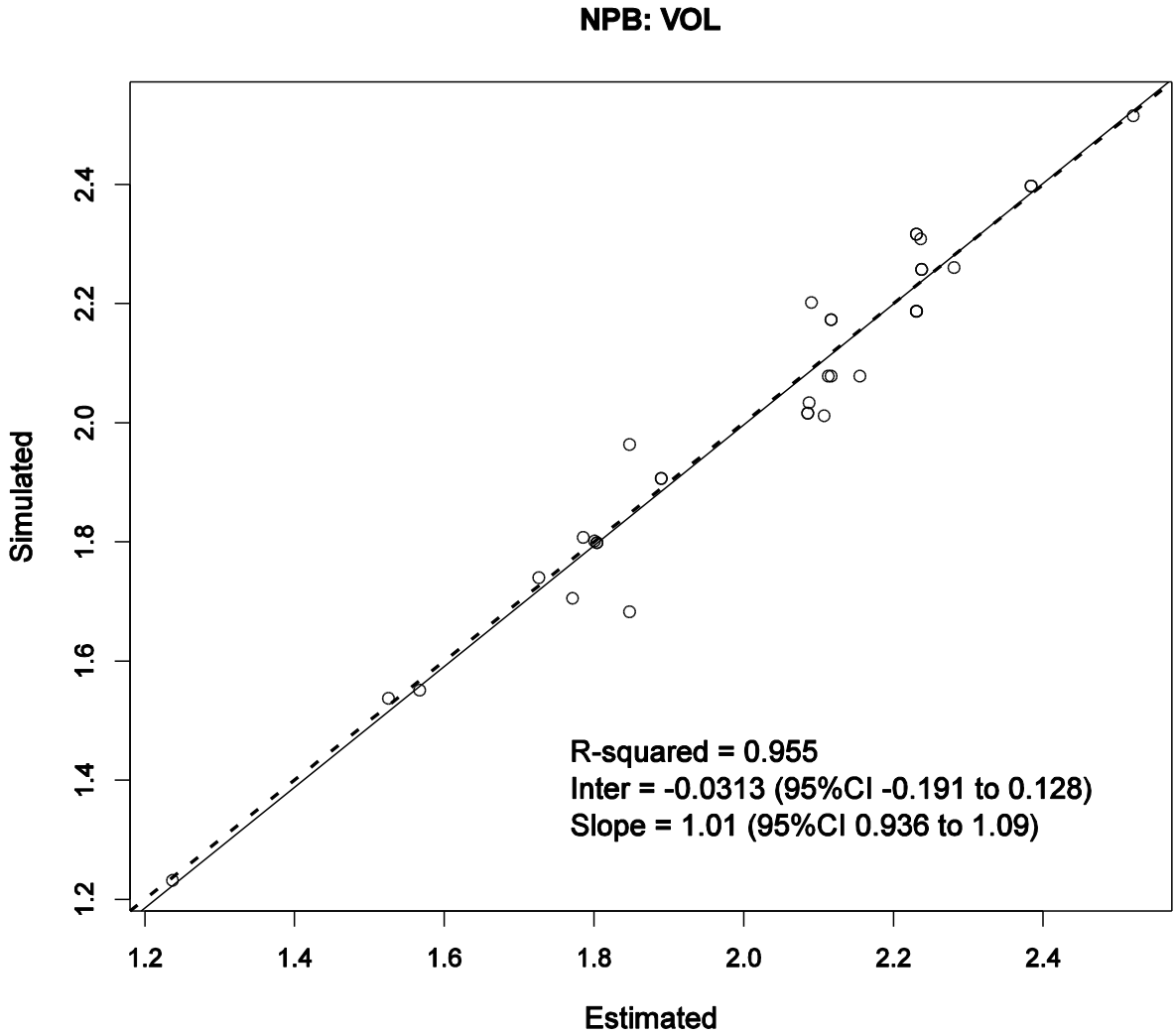

Figure 5: NPB error in true - fitted parameter values (KEL). Linear regression of simulated elimination constant vs. predicted elimination constant for each of the 35 simulated subjects using NPB algorithm.

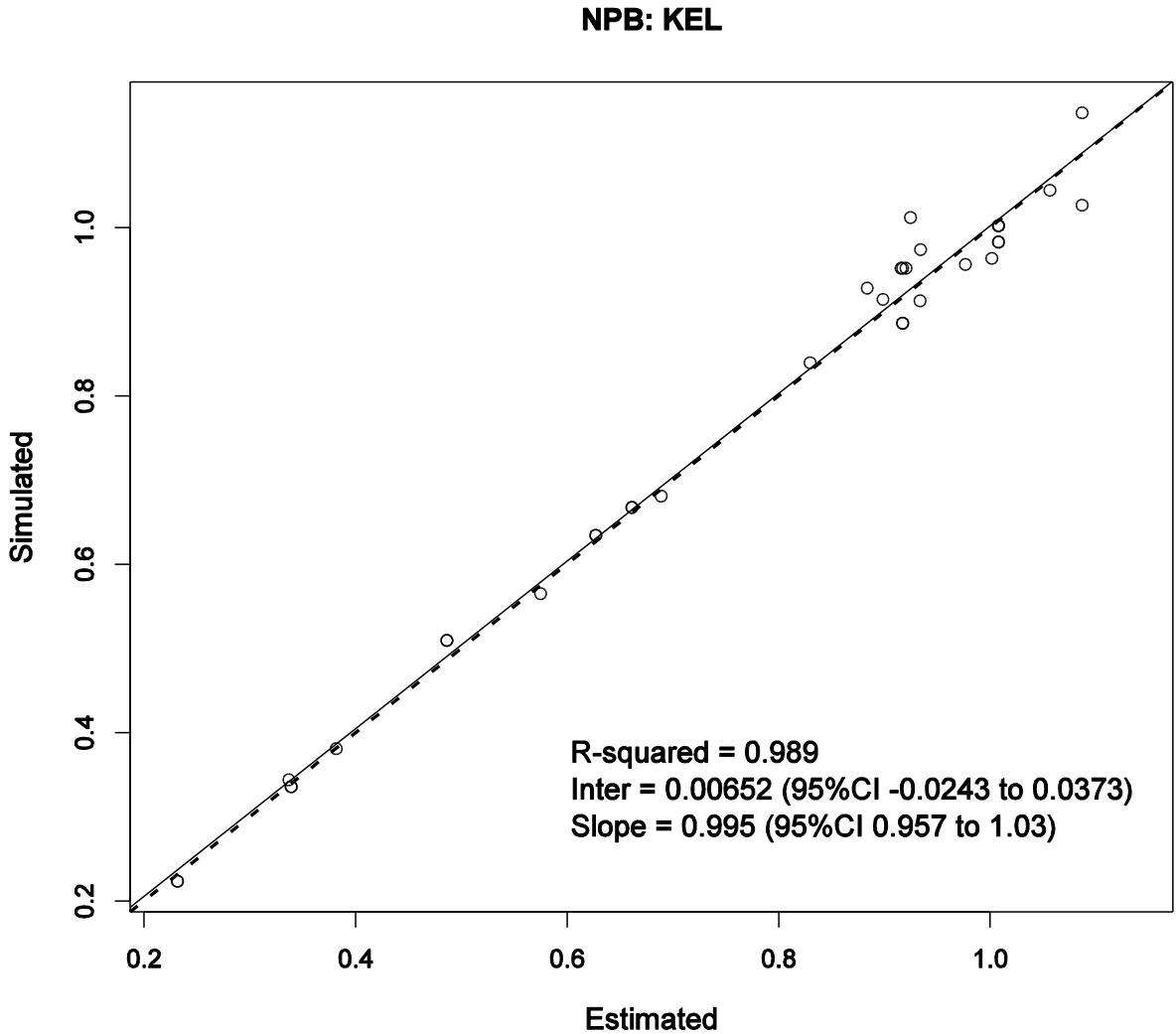

Figure 6: NPAG distribution and simulated values for volume of distribution (VOL). Distribution of simulated (true) parameter values are shown in black circles. The posterior distribution is represented by dark grey bars. True population distribution is shown as a dashed line.

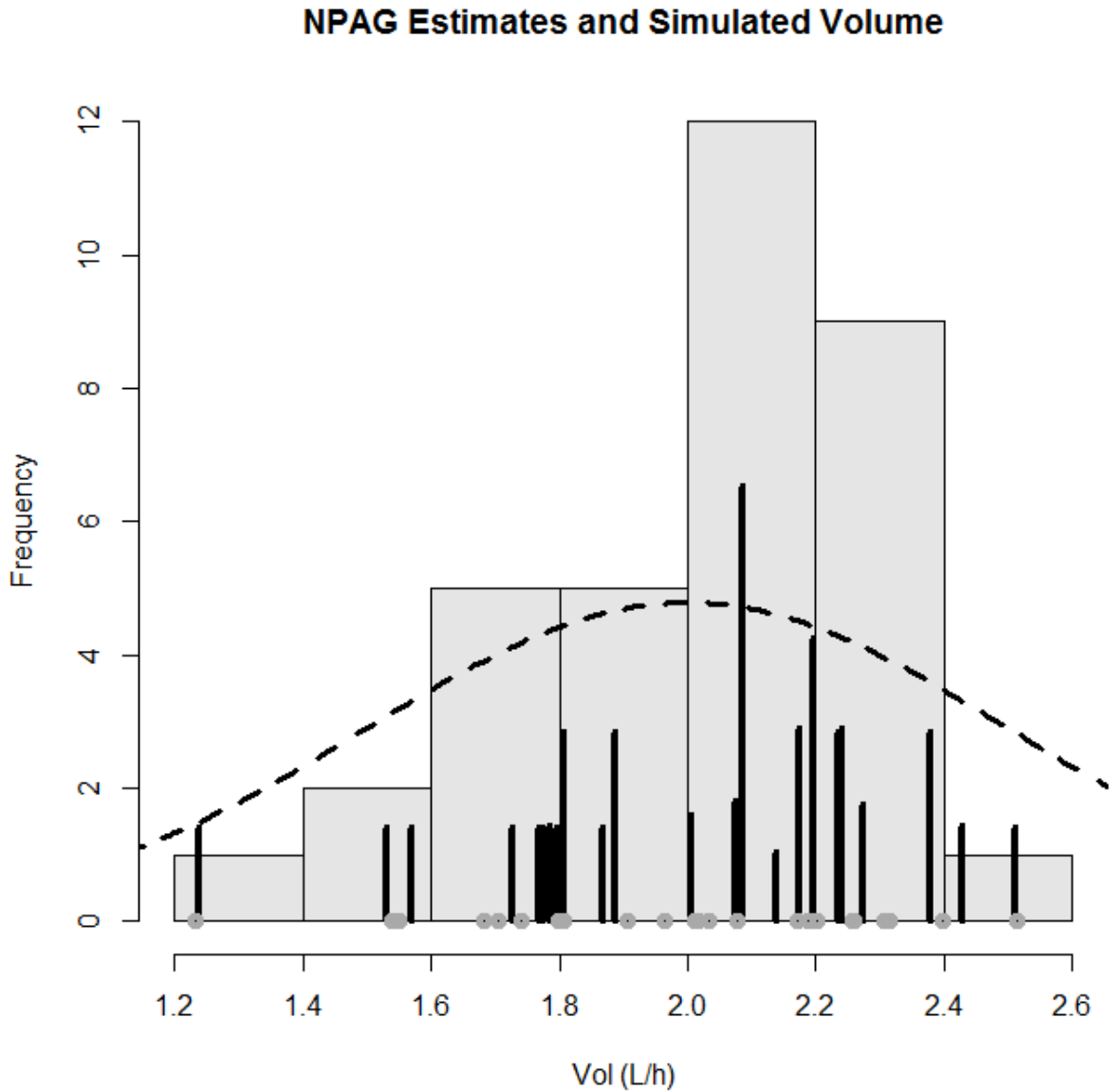

Figure 7: NPAG distribution and simulated values for elimination rate constant (KEL). Distribution of simulated (true) parameter values are shown in black circles. The posterior distribution is represented by dark grey bars. True population distribution is shown as a dashed line.

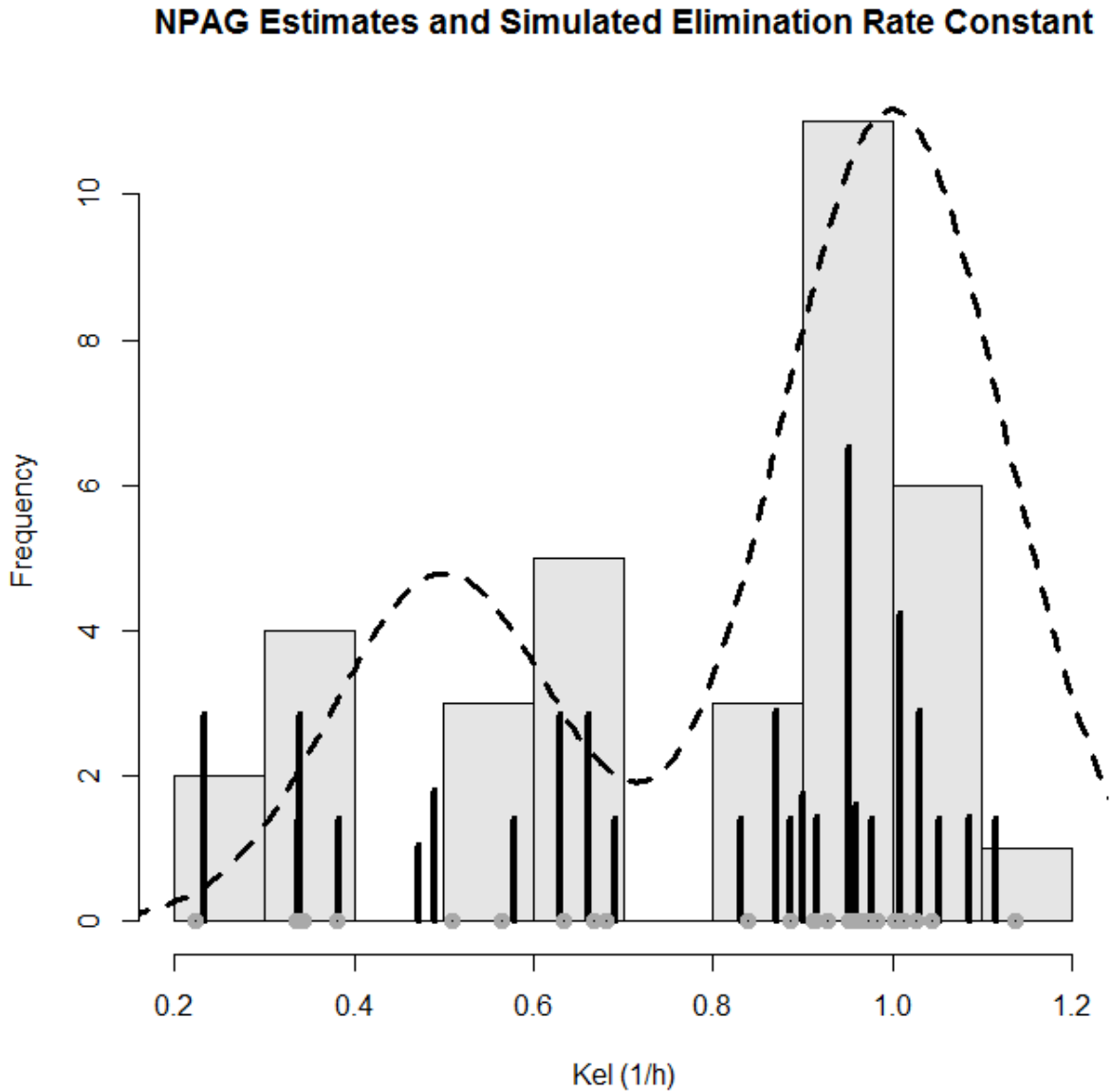

Figure 8: NPAG error in true - fitted parameter values (VOL). Linear regression of simulated volume of distribution vs. predicted volume of distribution for each of the 35 simulated subjects using NPAG algorithm.

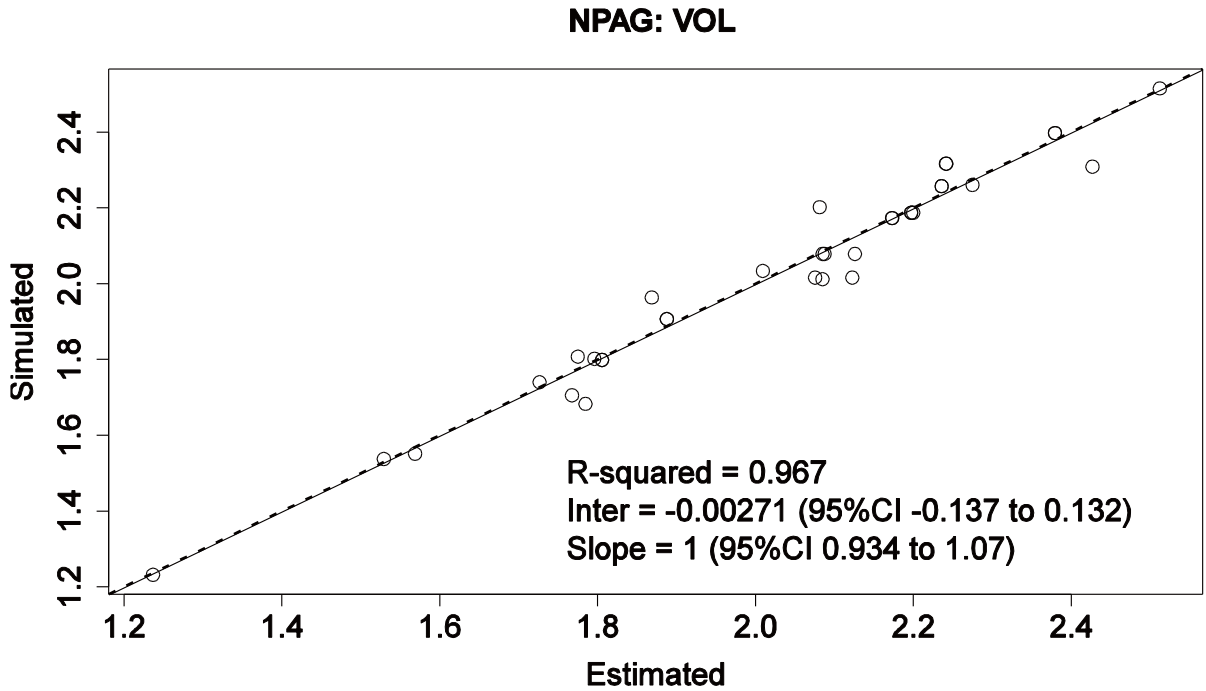

Figure 9: NPAG error in true - fitted parameter values (KEL). Linear regression of simulated elimination constant vs. predicted elimination constant for each of the 35 simulated subjects using NPAG algorithm.

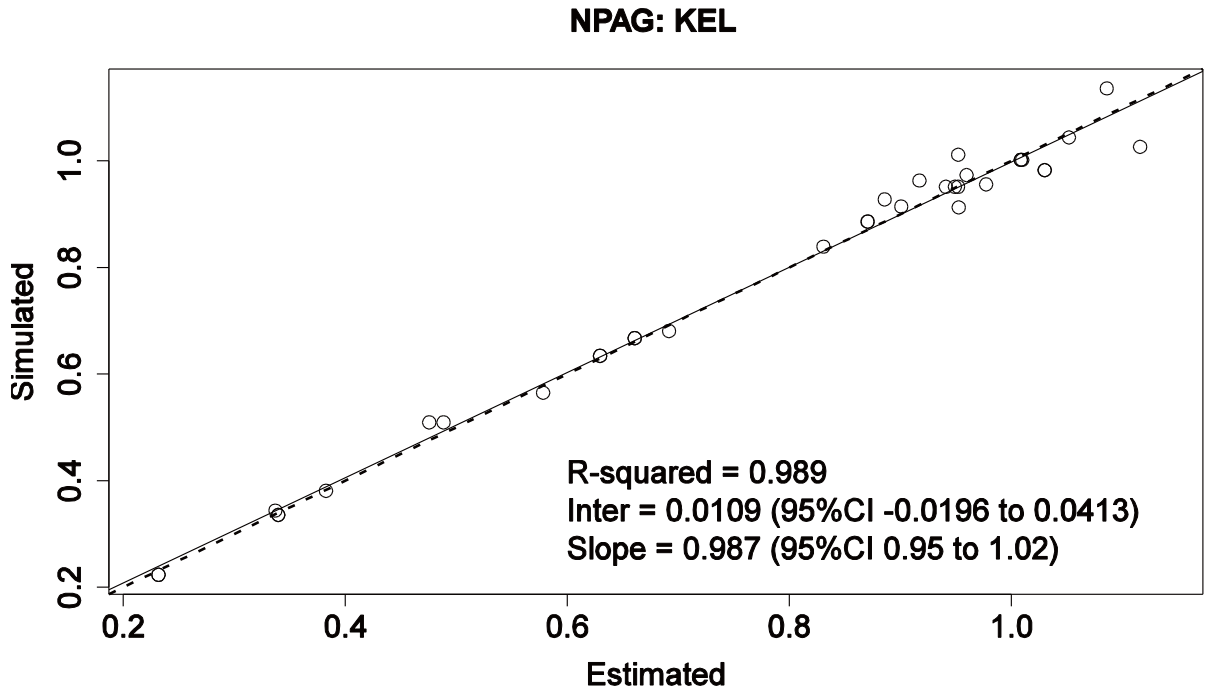